\documentclass{PoS}

\newcommand\cyg{Cyg~X-1}
\newcommand\gx{GX339$-$4}
\newcommand\fu{4U~1957+11}
\newcommand\rxte{\textsl{RXTE}}
\newcommand\chandra{\textsl{Chandra}}
\newcommand\meg{\textsl{MEG}}
\newcommand\pca{\textsl{PCA}}
\newcommand\hexte{\textsl{HEXTE}}
\newcommand\xmm{\textsl{XMM-Newton}}
\newcommand\as{$a^*$}
\newcommand\fc{f$_{\rm col}$}

\title{The Galactic Black Holes Cyg X-1, GX~339$-$4, and 4U~1957+11: In
Transition and at High Resolution}

\ShortTitle{Cyg X-1, GX339$-$4, and 4U~1957+11}

\author{\speaker{Michael A. Nowak}\\
        Massachusetts Institute of Technology, Kavli Institute for Astrophysics and Space Research\\
        E-mail: \email{mnowak@space.mit.edu}}

\abstract{I review here some of the open questions regarding the
geometry and emission mechanisms of galactic black hole candidates.
For hard states, I concentrate on the perspective of ``disk+Compton
coronae'' models (for discussions of jet models, see the papers by
Sera Markoff).  Specifically, I discuss the implications from our 10
year long \rxte\ monitoring campaign of \cyg.  I then present
simultaneous \rxte/\chandra\ observations of the ``soft state'' black
hole candidate \fu, and discuss to what extent it does or does not
allow one to test ``relativistic disk models''. The use of such models
has been claimed to measure black hole spin parameters.  I then
briefly present a particularly freaky-weird observation of \gx, where
the source ``fell off'' the usual radio/X-ray correlation in the
low/hard state.  Questions addressed by the above observations
include: are the Compton corona models unique fits to the data?
(No. Jets work equally well, and simple broken power laws work better
still. We argue that the latter models indicate multiple, broad-band
continuum components.) Is there good evidence for a receding disk as
sources transit into the hard state?  (The jury is still out.) What
does the relativistically broadened Fe line tell us? (Sometimes the
disk, \emph{even into quiescence}, stays very close to the central
object, in contrast to expectations of ADAF models.)  How much
better/more necessary are recently discussed relativistic disk models?
(I am very doubtful that such models will ever usefully measure black
hole spin.)}

\FullConference{VI Microquasar Workshop: Microquasars and Beyond\\
		September 18-22 2006\\
		Societ\`a del Casino, Como, Italy}

\begin{document}

\section{Introduction}

As of this, the sixth conference in this series, I believe that most
of us are comfortable with the notion that \emph{all} black hole
candidates (BHC) are, have been, or potentially could be
\emph{microquasars}, i.e., galactic compact object sources that show
significant jet activity (usually in the radio).  The main question
that determines whether or not BHC exhibit steady jet activity is
whether or not they are in the spectrally hard (i.e., ``low'') state
\cite{fender:99b,corbel:00a}.  (For an unusual quasi-exception to
this, the reader is referred to the observation of \gx\ discussed at
the end of this article.)  The questions many of us are now focused on
are: What are the emission mechanisms responsible for BHC hard states
(corona and/or jet)?  What is the hard state geometry?  (The latter is
especially relevant to coronal models of hard states.)  Is rapid black
hole spin required for launching a jet?  (As the \emph{Fender
Conjecture} states: if jets require rapid spin, then all black holes
are rapidly spinning, since all seem to exhibit jets in their hard
state.)  Can we measure black hole spin?

In the following, I address some of these issues via a series of
pointed monitoring observations of three BHC: \cyg, \fu, and \gx.
We have been monitoring \cyg\ with \rxte\ for nearly 10 years now, and
what follows are some highlights from our work which can be read in
detail elsewhere
\cite{dove:98a,nowak:99a,pottschmidt:02a,gleissner:04a,gleissner:04b,wilms:06a}
that are particularly relevant to Comptonization and jet models.
Next, I discuss recent \chandra\ and \rxte\ observations of the soft
state BHC, \fu\ (Nowak, et al., in prep.).  As this source is
persistently soft, with minimal hard tail emission
\cite{nowak:99d,wijnands:02c}, it becomes an excellent test of notions
of spectrally measuring spin (despite, as we shall discuss, lack of
knowledge of its distance or mass).  Finally, I will briefly show an
unusual, radio under-luminous, hard state of \gx\ (Nowak et al., in
prep.).

Throughout, with one exception, all plotted spectra are shown as
``flux-corrected'' spectra (i.e., adjusted \emph{solely} based upon
the detector {\tt arf} and {\tt rmf}, with \emph{no} reference to the
model), as calculated with the \textsl{ISIS}\footnote{\emph{All} the
models of \textsl{XSPEC}, with most of the programmability of
\textsl{IDL} or \textsl{MATLAB}, plus arbitrarily and easily
extendible with almost any \textsl{Fortran}, \textsl{C}, or
\textsl{C++} library, plus parallel processing via the \textsl{PVM}
module, plus passing of data back and forth to \textsl{DS9} via the
\textsl{XPA} module, plus $\ldots$.  Oh, hell, if you're still using
\textsl{XSPEC} instead, you deserve what you get.  For an
introduction, see {\tt
http://space.mit.edu/home/mnowak/isis\_vs\_xspec/index.html}.}
analysis system \cite{houck:00a}.  As such, and in contrast to
\textsl{XSPEC} ``unfolded'' spectra, you will not see \emph{assumed}
model features \emph{falsely} mirrored in the data.  The data are what
they are, and any observed structure in the ``flux-corrected'' spectra
are indicative of physical reality and/or response matrix features
(not all of which are necessarily properly characterized).

\section{\cyg}

\begin{figure}
\begin{center}
\includegraphics[width=0.44\textwidth]{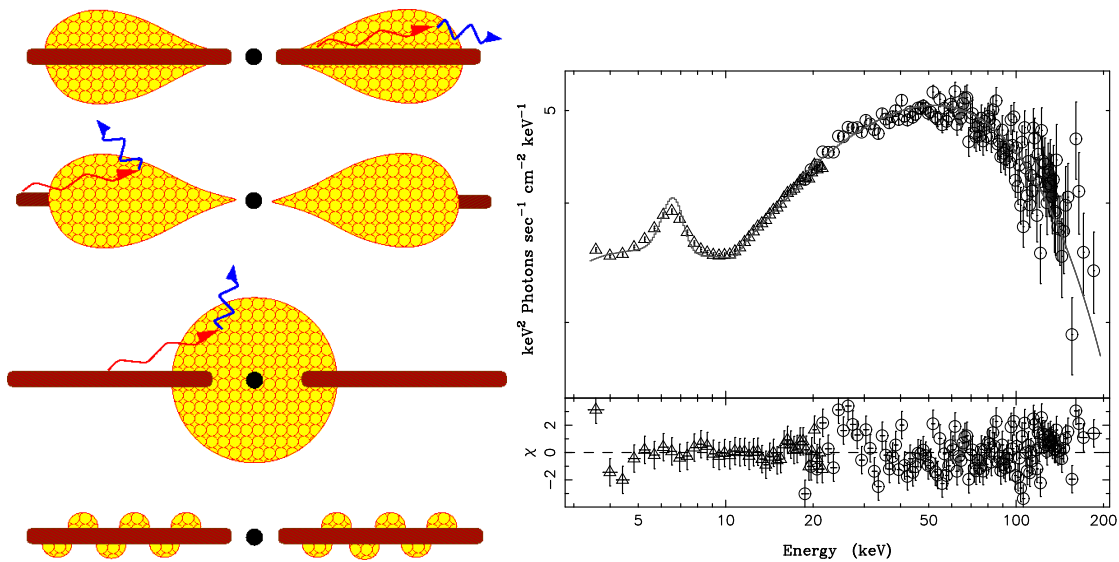}
\includegraphics[width=0.53\textwidth]{ir_co_cyg.ps}
\end{center}
\caption{Left: A number of suggested coronal geometries.  The middle
two are currently very popular, e.g., in ADAF models; however, their
revival in Compton codes was somewhat independent of ADAFs (see text).
Right: `Flux-corrected' (i.e., \emph{not} following \textsl{XSPEC}'s
unfolding scheme - see text and Nowak et al. 2005) \rxte\ spectra of a
very hard state of \cyg.}
\label{fig:coronae}
\end{figure}

\begin{figure}
\begin{center}
\includegraphics[width=0.48\textwidth]{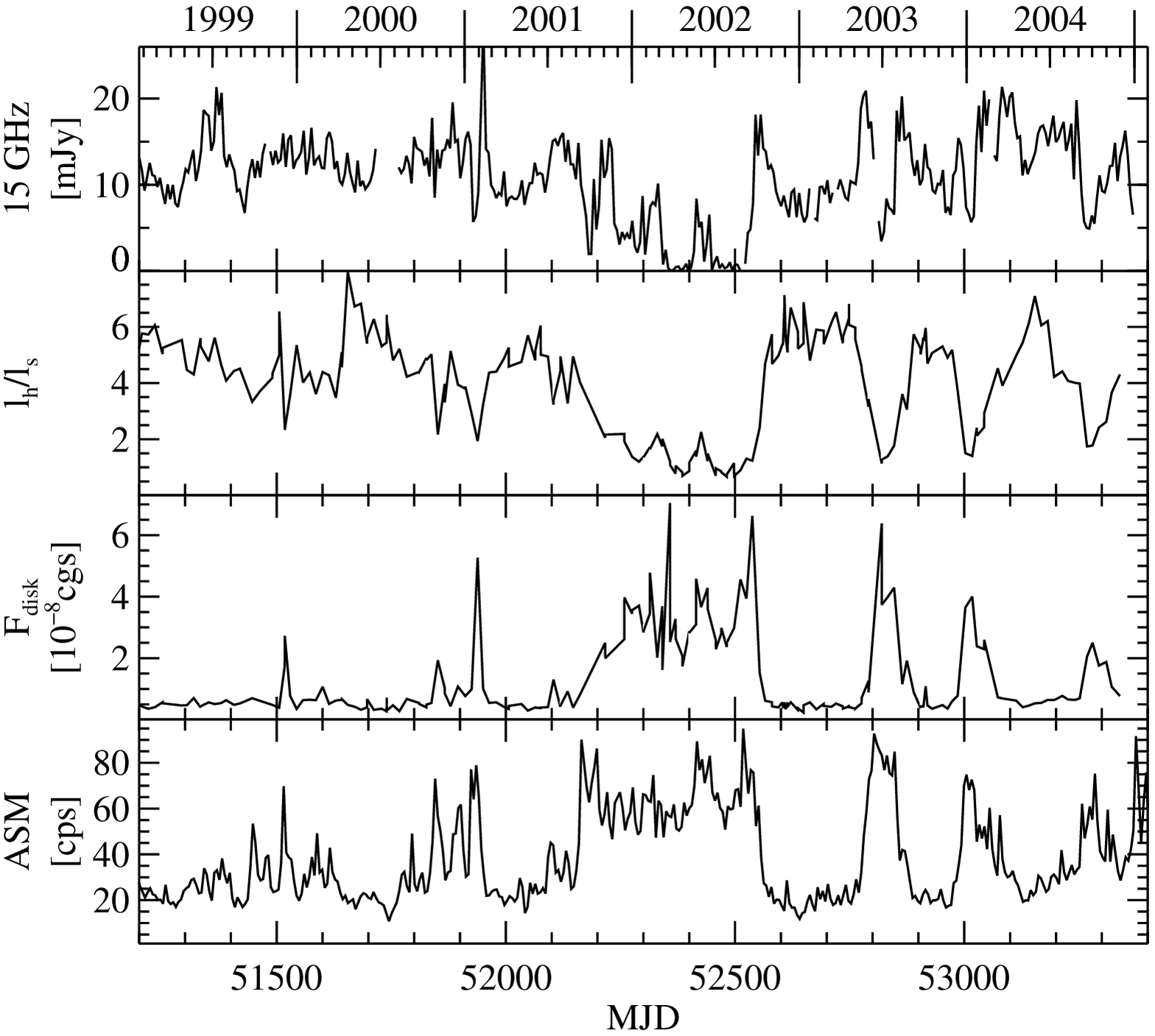}
\includegraphics[width=0.48\textwidth]{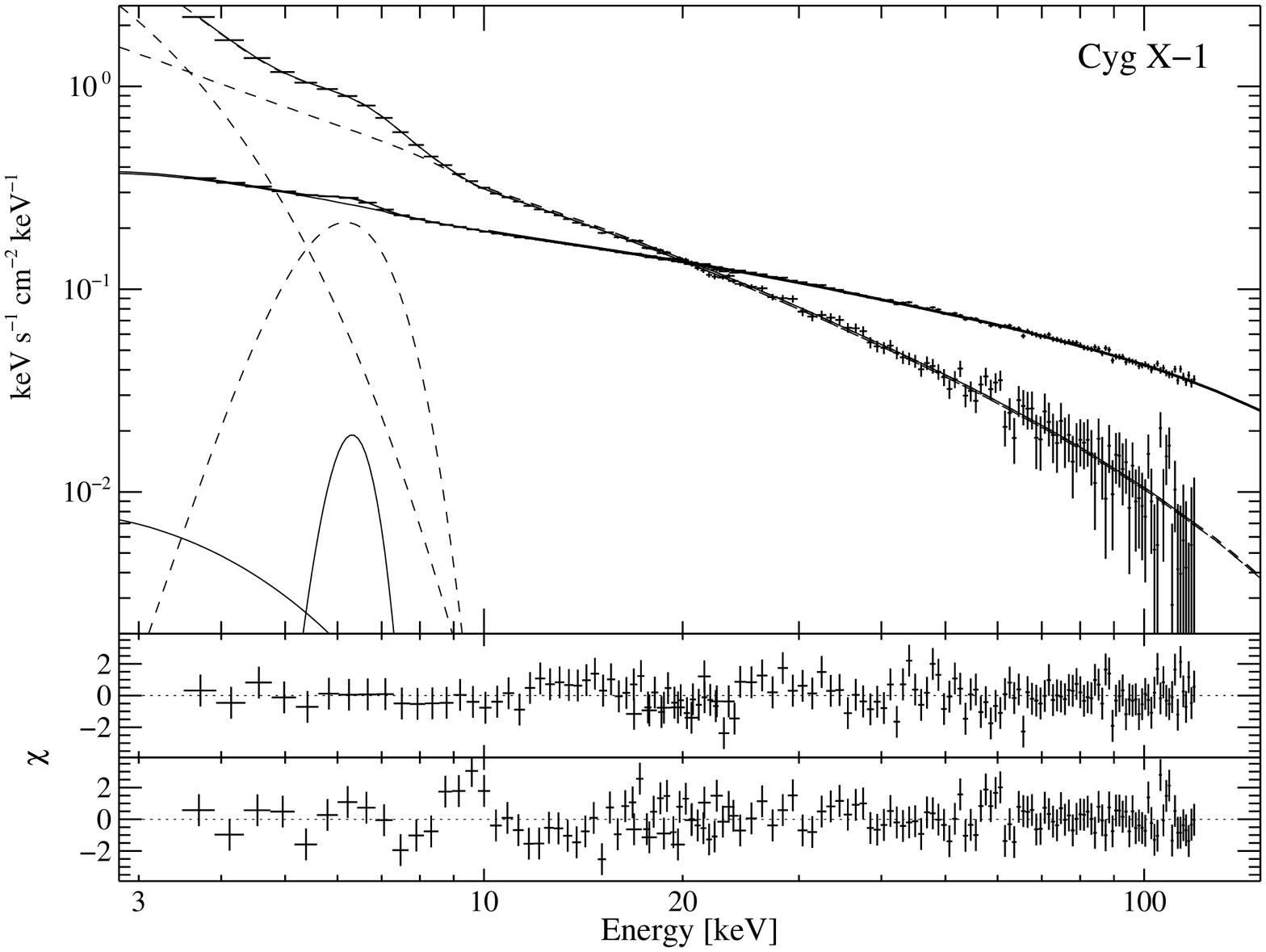}
\end{center}
\caption{Left: \textsl{ASM} flux, radio flux, coronal compactness and
disk flux (the latter two from {\tt phabs*(diskpn+eqpair+gauss)} fits)
from our pointed \rxte\ observations of \cyg. Right: Our softest and
hardest \cyg\ observations. (Sadly, the latter shown as the potentially
misleading \textsl{XSPEC} unfolded spectra. The individual {\tt
diskpn}, {\tt eqpair}, and {\tt gauss} model components are also
shown.)}
\label{fig:extreme}
\end{figure}

\cyg\ is perhaps the most famous and one of the best studied of the
BHC.  A review of the overall system properties can be found in
\cite{nowak:99a}. \cyg\ served as one of the original motivators for
the definition of the ``hard'' (i.e., ``low'') and ``soft'' (i.e.,
``high'') states, despite two tremendous drawbacks.  First, being at
least partially wind-fed rather than strictly Roche lobe-fed (i.e.,
``focused-wind accretion''), it likely has something of a truncated
accretion disk, different than the typical X-ray nova accretion disk.
Second, \cyg\ is perhaps one of the worst examples of ``soft state''
transitions.  As we will discuss, the \cyg\ state transitions are more
of degree than kind (see Fig.~\ref{fig:extreme}), with \cyg\
\emph{never} entering a ``disk dominated'' state.

A typical \cyg\ hard state spectrum, as observed by \rxte\ in the
3--200\,keV band, is shown in Fig.~\ref{fig:coronae}.  The salient
features -- present in \emph{all} \rxte\ spectra of \cyg\ -- are a low
energy (3--10\,keV) power law that hardens above 10\,keV, a (typically)
broad line feature near 6.4\,keV, and an exponential rollover at high
energies (usually $>20$\,keV).  \cyg\ ``soft states'' also often require
an additional soft component (which can be modeled with a
simple phenomenological disk component).  In fact, the simple
phenomenological model as described above fits \rxte\ data
\emph{better than any sophisticated Comptonization or jet model that
we have tried} \cite{wilms:06a}.  As a corollary, \emph{including}
parameterizing absorption and a relative \pca/\hexte\ normalization
constant, any model that uses more than 11 free parameters to
characterize the \rxte\ data is likely over-parameterized.

Comptonization of soft (seed) disk photons in a hot ($\approx
100$\,keV) corona with $\tau_{es}$ of order a few has long been
proposed as a physical description of BHC spectra
\cite{eardley:75a,sunyaev:79a}.  Debate centers around the geometry,
with `sphere+disk' models currently being very popular (see
Fig.~\ref{fig:coronae}).  The `revival' of the `sphere+disk' geometry
for models of \cyg\ in fact occurred independently of Advection
Dominated Accretion Flow (ADAF) models, which also posit this
geometry.  It had been noted that unless the corona is `photon
starved' (i.e., only a fraction of seed photons enter the corona), it
is simply too difficult to achieve temperatures high enough to produce
spectra as hard as those seen in \cyg\ \cite{dove:97b,poutanen:97b}.
`Pill box' geometries (i.e., the bottom of Fig.~\ref{fig:coronae}),
although photon starved, produce too much reflection, \emph{unless}
some process like relativistic beaming \cite{beloborodov:99a} is also
invoked.

In \cite{wilms:06a}, we successfully fit the \rxte\ \cyg\ spectra with
the {\tt eqpair} model of Coppi \cite{coppi:99a}, if we include
additional disk and line components (smeared reflection is included in
the {\tt eqpair} model).  Fits to the hardest and softest spectra are
shown in Fig.~\ref{fig:extreme}. The relevant coronal fit parameters
then become the \emph{relative} compactness (i.e., energy divided by
radius) of the corona to that of the disk, and the coronal optical
depth.  Other Comptonization models also work (with, for example,
Compton $y$ parameter taking the role of relative compactness).
Overall \emph{trends} remain unchanged among these models, especially
on ``broad'' features such as fluxes; however, \emph{absolute} numbers
(e.g., reflection fraction, Fe line strength \emph{and} width) are
altered, often in systematic ways \cite{wilms:06a}.  Trends from the
{\tt eqpair} fits are shown in Fig.~\ref{fig:extreme}.

\begin{figure}
\begin{center}
\includegraphics[width=0.32\textwidth]{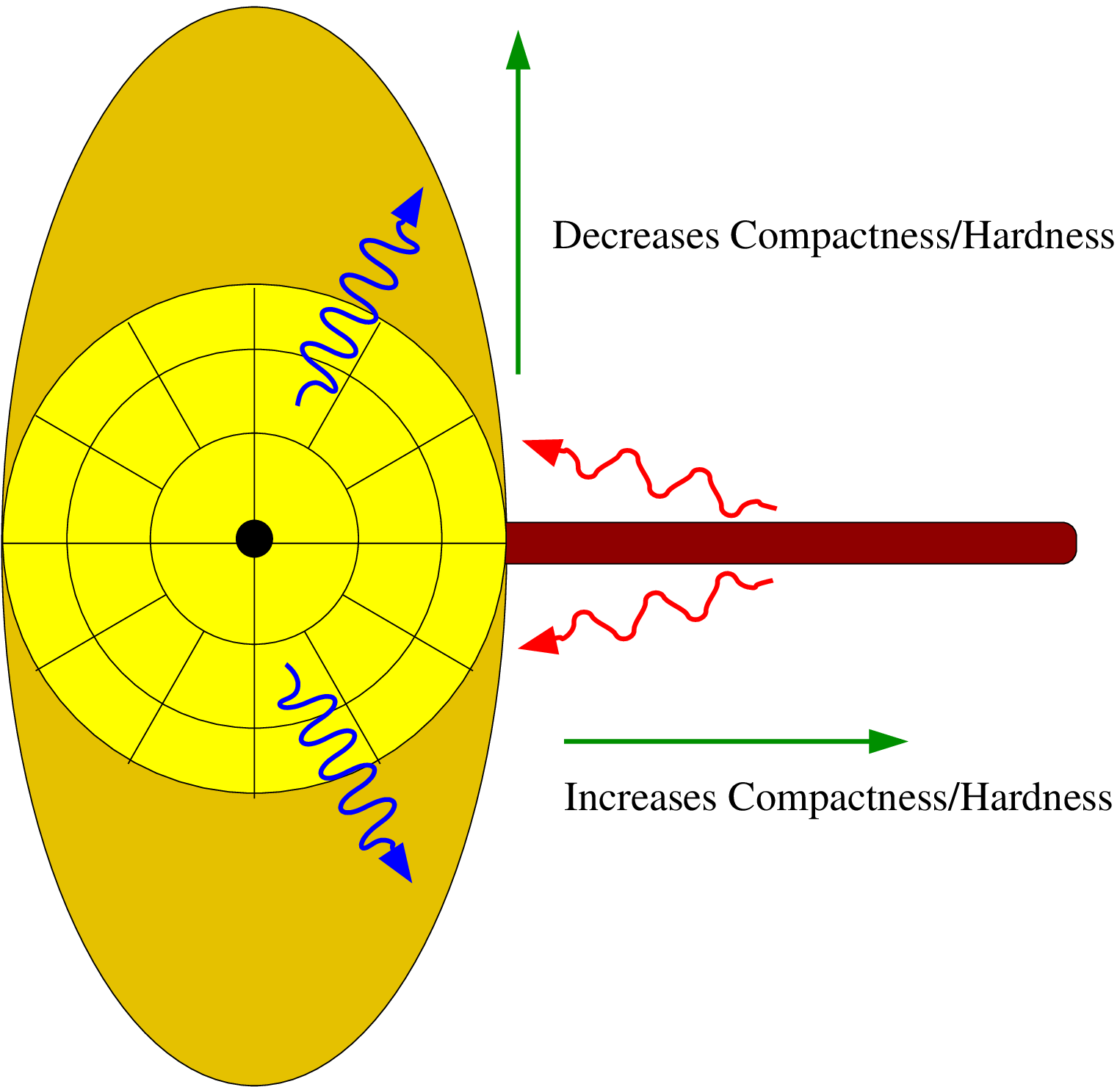}
\includegraphics[width=0.32\textwidth]{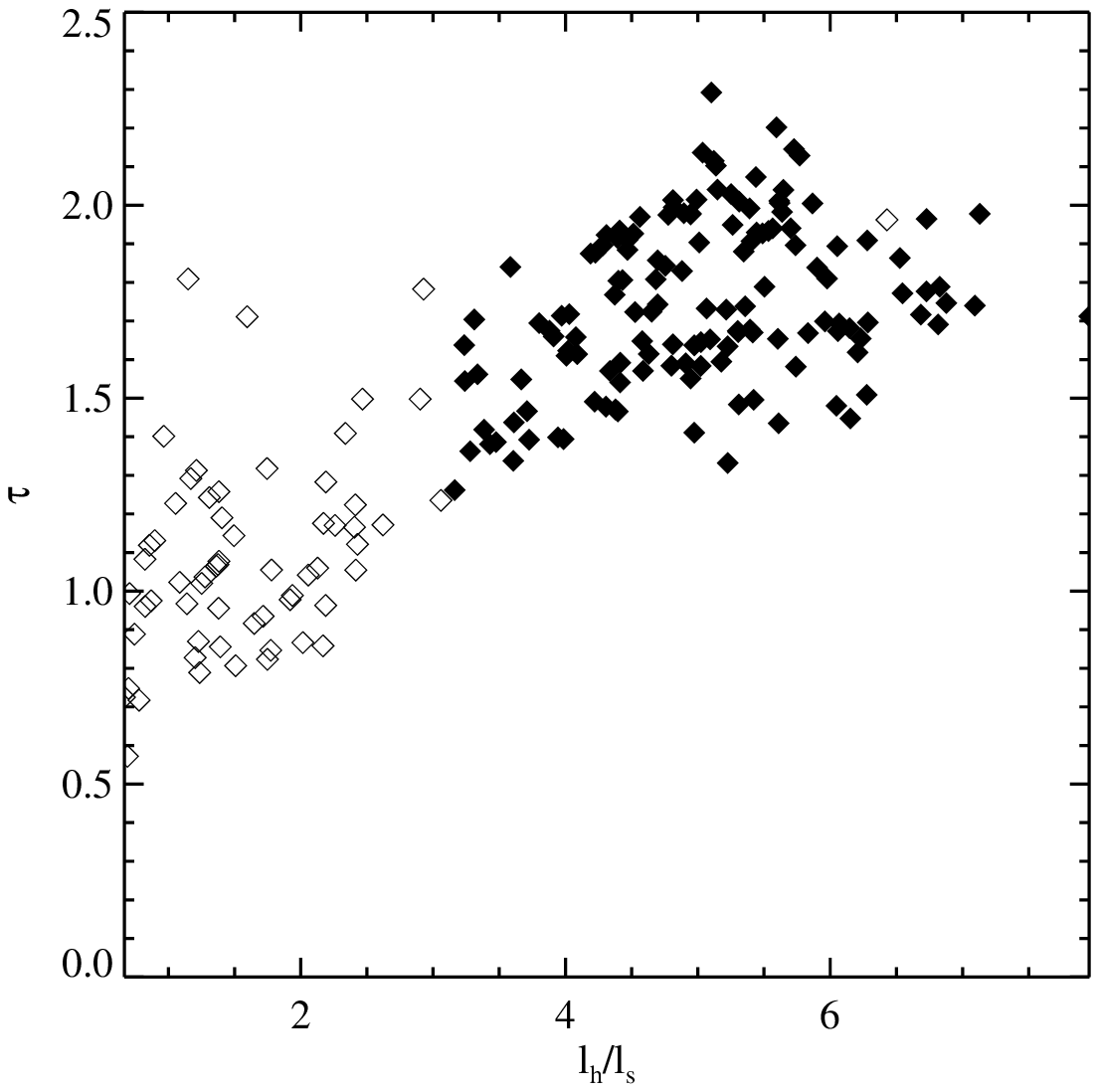}
\includegraphics[width=0.32\textwidth]{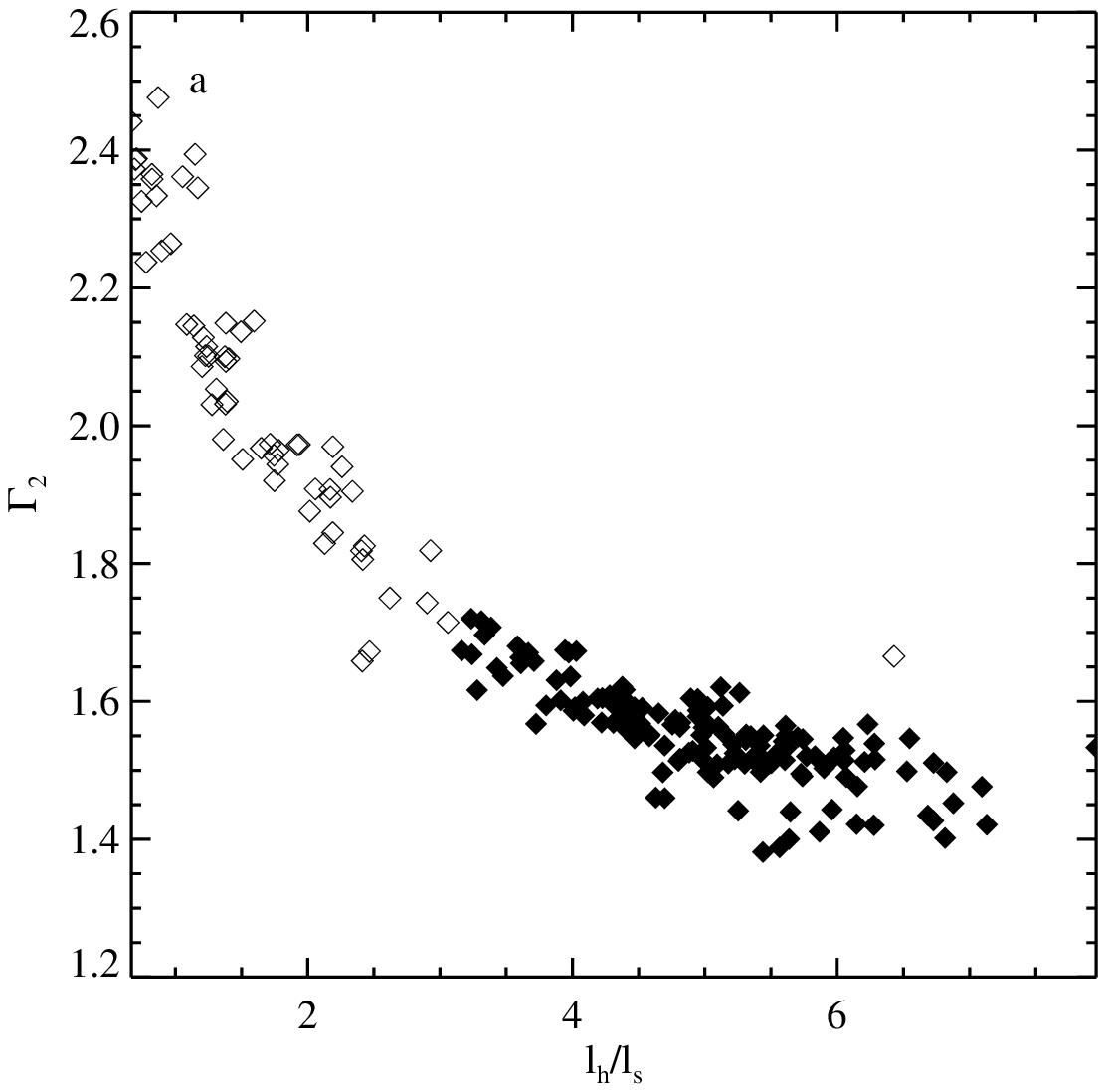}
\end{center}
\caption{Left: Schematic of the sphere+disk coronal geometry, with
expectations for how the compactness/hardness could change with simple
changes of the geometric parameters.  Middle: coronal optical depth
vs. compactness ratio for our pointed observations of \cyg.  Here, and
throughout, clear points will refer to the Remillard \& McClintock
\protect{\cite{remillard:06a}} power law photon index-based definition
of the soft state ($\Gamma_1 > 2.2$), while the solid points refer to
the photon index-based definition of the hard state.  Right: higher
energy power law photon index ($\Gamma_2$) vs. ratio of coronal
compactness.  }
\label{fig:ntdisk}
\end{figure}

\begin{figure}
\begin{center}
\includegraphics[width=0.32\textwidth]{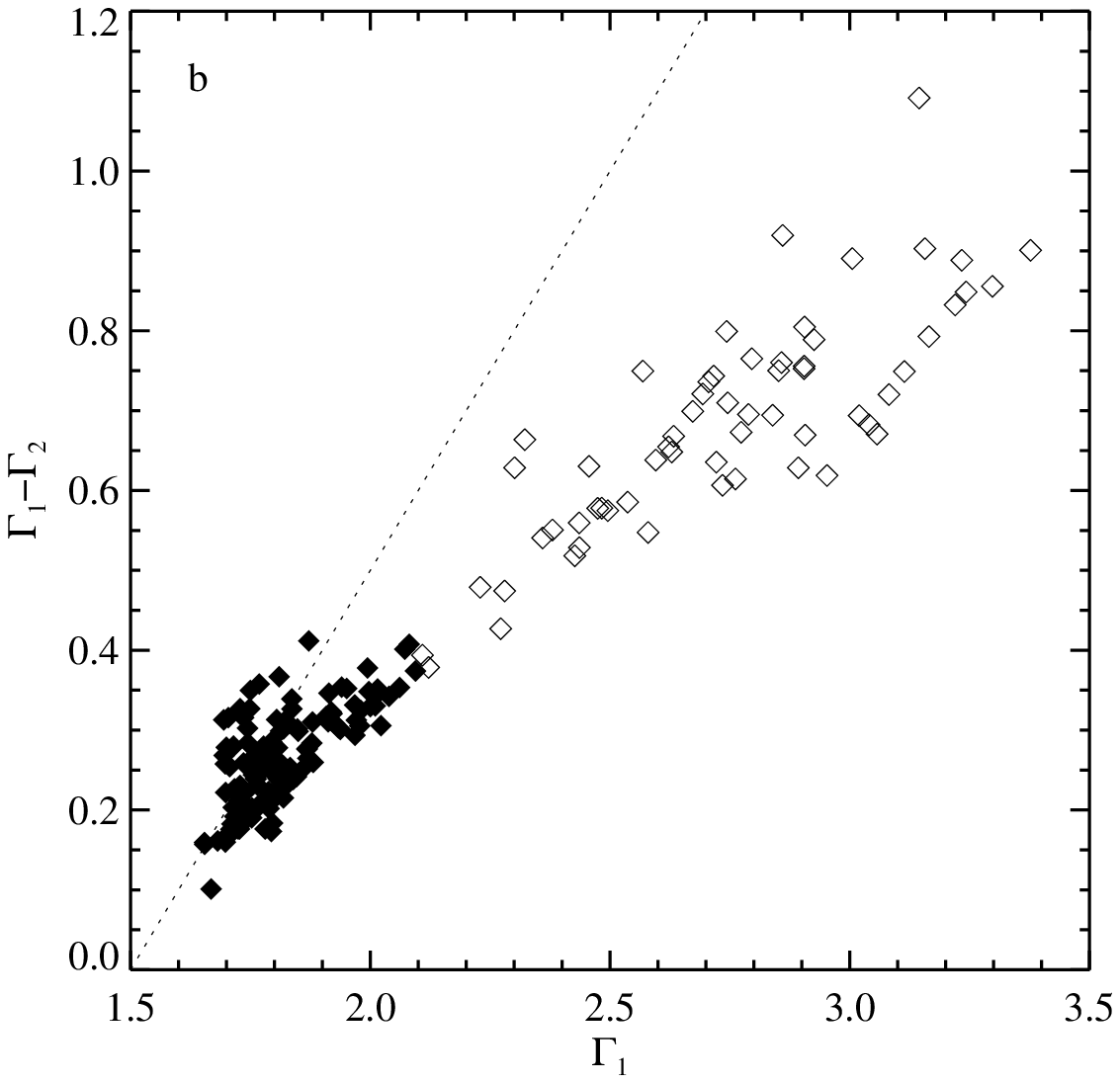}
\includegraphics[width=0.32\textwidth]{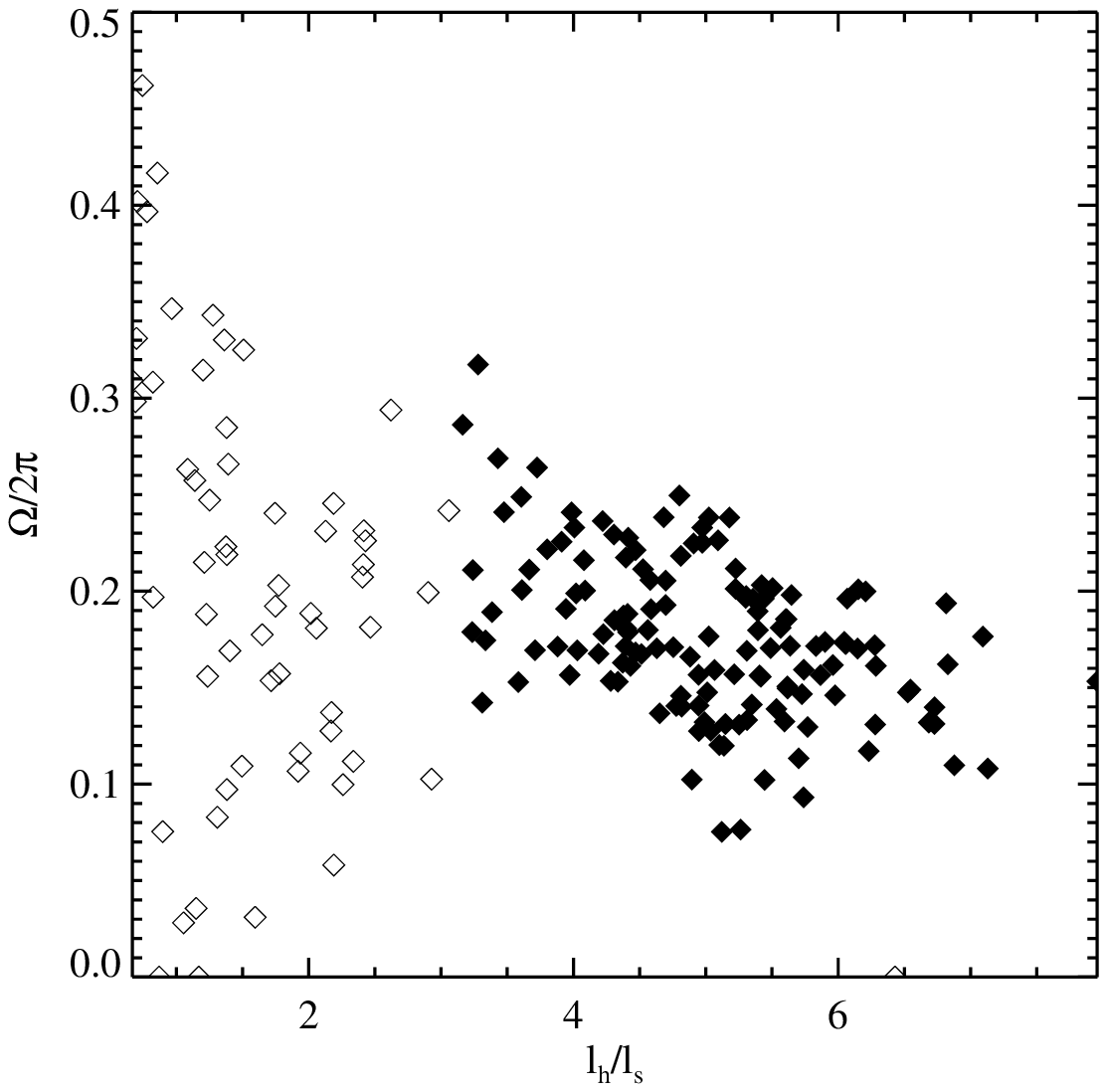}
\includegraphics[width=0.32\textwidth]{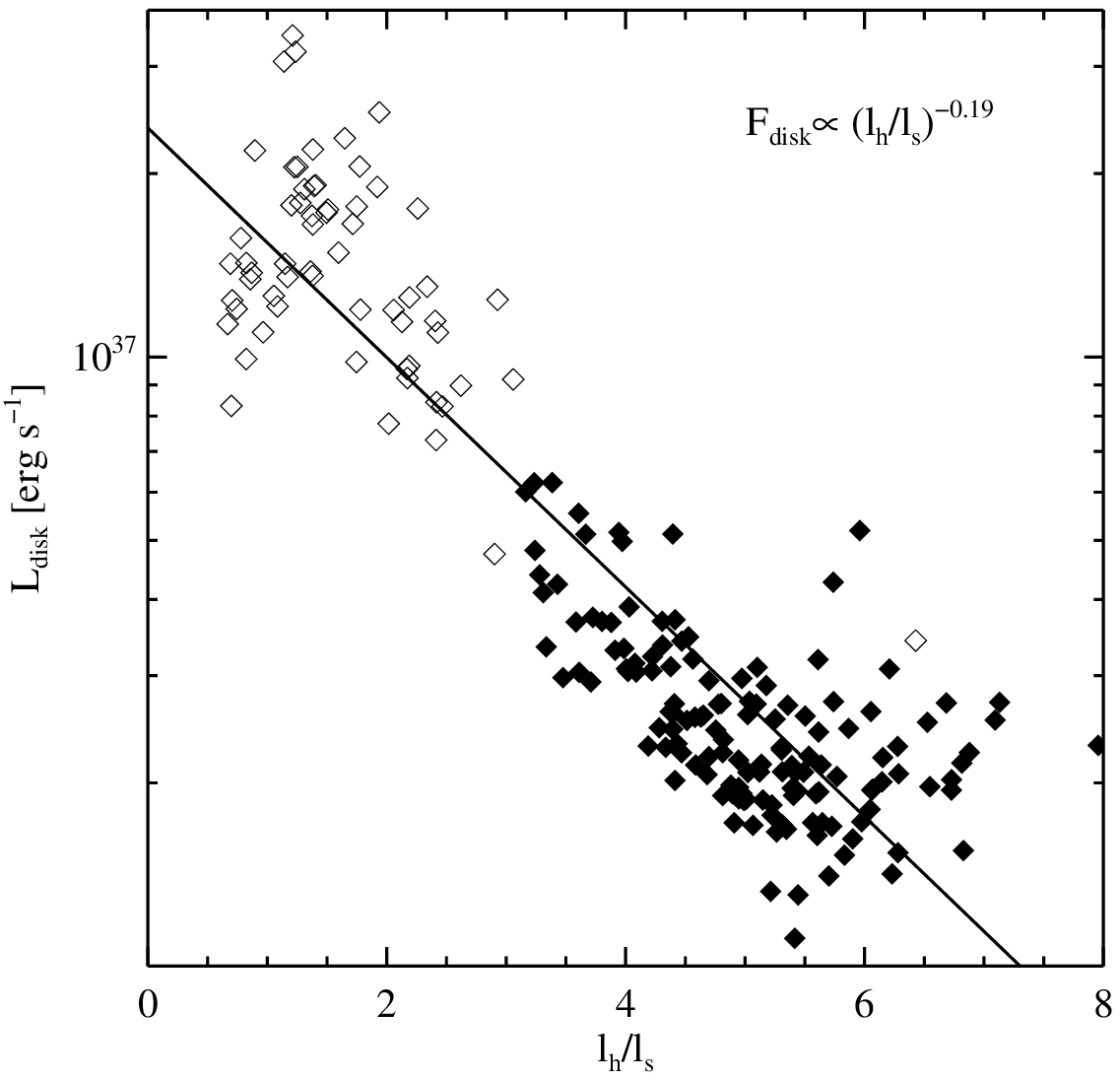}
\end{center}
\caption{Left: Difference in photon indices vs. lower energy photon
index for broken power law fits to \cyg. Middle: Reflection fraction
vs. compactness ratio for coronal model fits.  Right: Disk flux
vs. compactness ratio for coronal model fits.}
\label{fig:eqcors}
\end{figure}

Using the {\tt eqpair} model in this way only roughly approximates the
sphere+disk geometry, but it does provide good fits to the data. There
is also a nearly one-to-one correspondence with broken power law fits
(see Fig.~\ref{fig:ntdisk}).  An important point to note here is that
what is a very clear and distinct power law break vs. power law index
correlation (Fig.~\ref{fig:eqcors}) is not \emph{solely} attributable
to a reflection fraction-hardness correlation (i.e.,
\cite{zdziarski:99a}) in Comptonization models.  Although a
\emph{weak} relation between reflection vs. coronal compactness (i.e.,
hardness) is seen, the $\Delta \Gamma-\Gamma_1$ correlation is
\emph{more dominated} by a disk flux-coronal compactness correlation
(Fig.~\ref{fig:eqcors}).  That is, the break at 10\,keV is largely
driven by the relation between \emph{two broad band continuum
components being observed within the \rxte\ band}.  To be clear, the
presence of the Fe line indicates that there \emph{must} be
reflection, but its exact value and its correlations cannot be
determined independently from the broad-band continuum components
assumed and fit in the 3--200\,keV regime.

Within the context of the sphere+disk Comptonization model, aside from
changing the energetics or optical depth of the corona, one can alter
the spectral hardness by changing the size of the corona and/or the
radius of the transition region between disk and corona (see
Fig.~\ref{fig:coronae}).  Such geometry changes have been inferred
from spectral fits, e.g., the reflection-hardness correlations, but as
we show above, that correlation is weak, is not truly
self-consistently calculated within the context of the fit model, and
it does depend upon the presumed Comptonization and reflection model
(i.e., {\tt reflect*comptt} models show a much more pronounced
correlation, which is likely systematic rather than physical in
nature; \cite{wilms:06a}).  Thus, researchers have turned to timing
features to search for further clues as to geometry changes in the
\cyg\ system.

\begin{figure}
\begin{center}
\includegraphics[width=0.5\textwidth,angle=270]{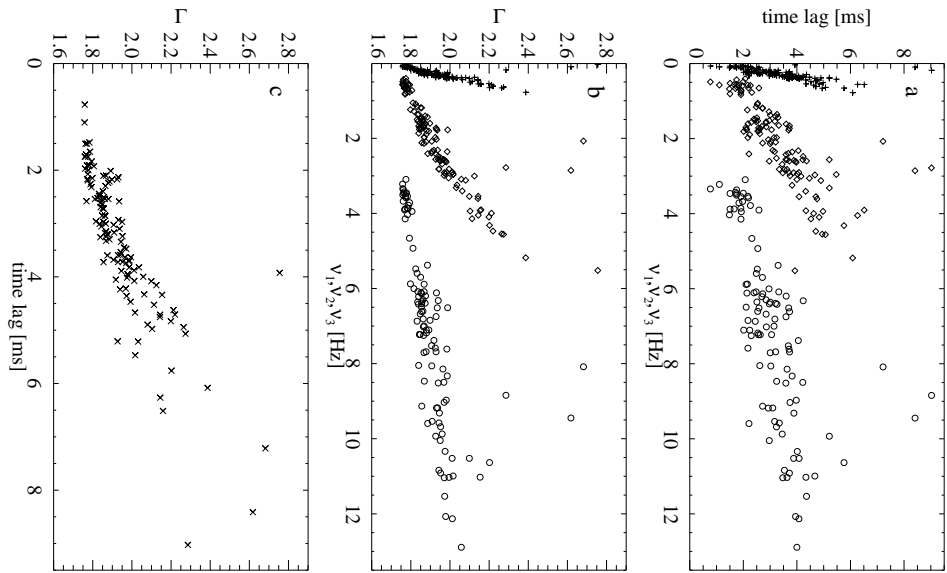}
\end{center}
\caption{The dependence of characteristic power spectral density (PSD)
frequencies upon photon index (from Pottschmidt et al. 2002; $\Gamma
\approx \Gamma_1$).  Also shown are time lags between hard and soft
variability vs. both photon index and characteristic PSD frequency.}
\label{fig:katja}
\end{figure}

\begin{figure}
\begin{center}
\includegraphics[width=0.27\textwidth]{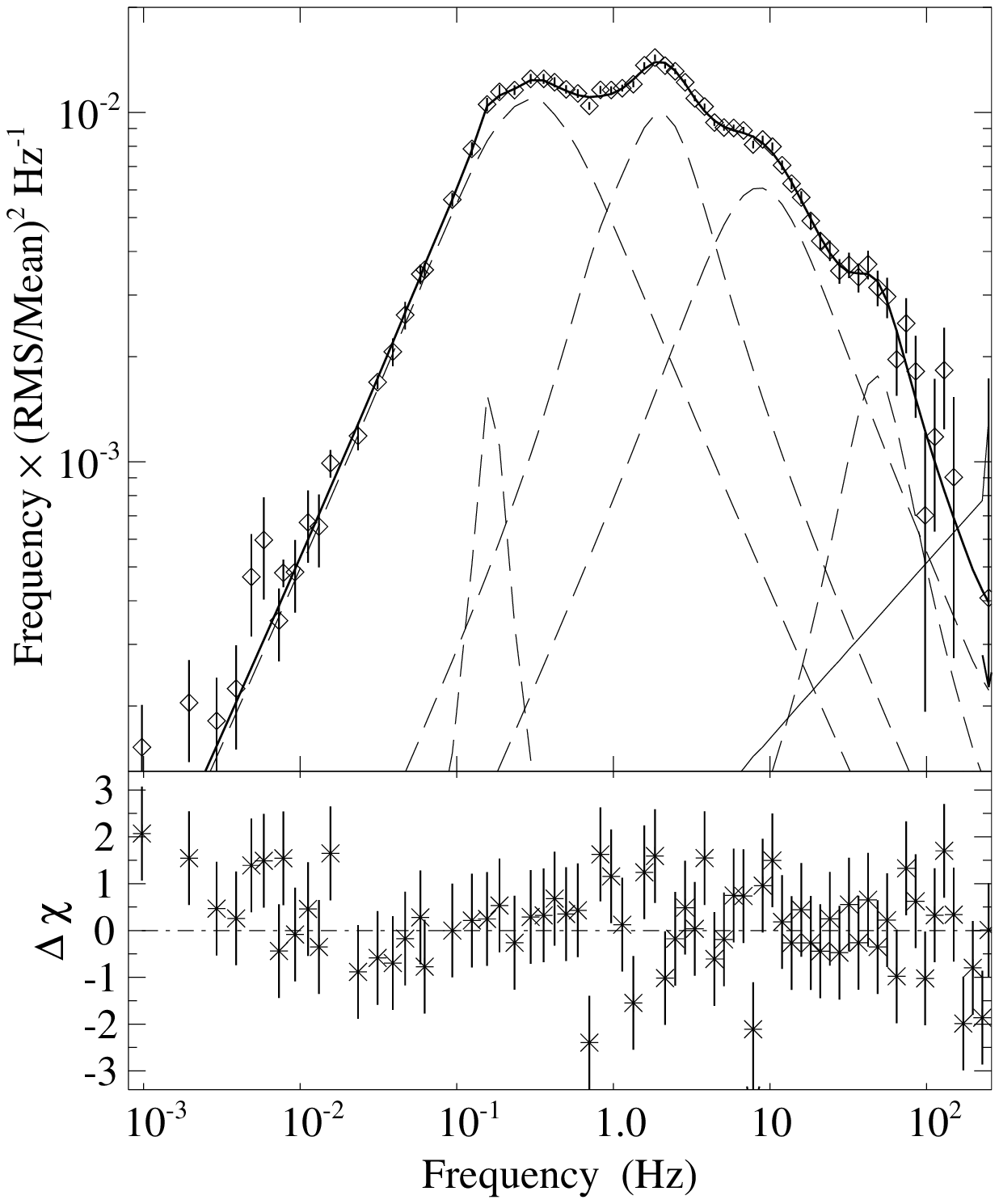}
\includegraphics[width=0.355\textwidth]{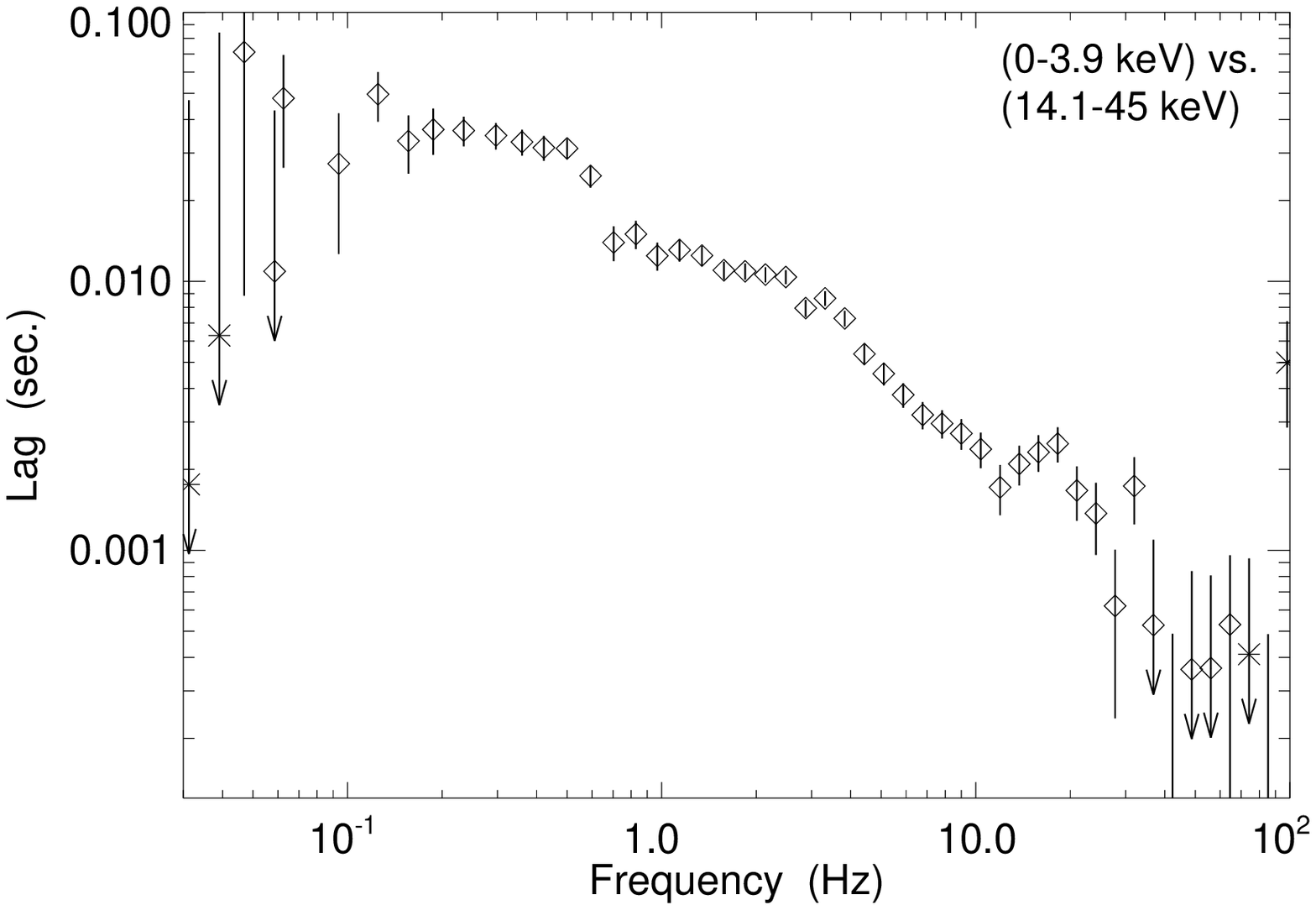}
\includegraphics[width=0.355\textwidth]{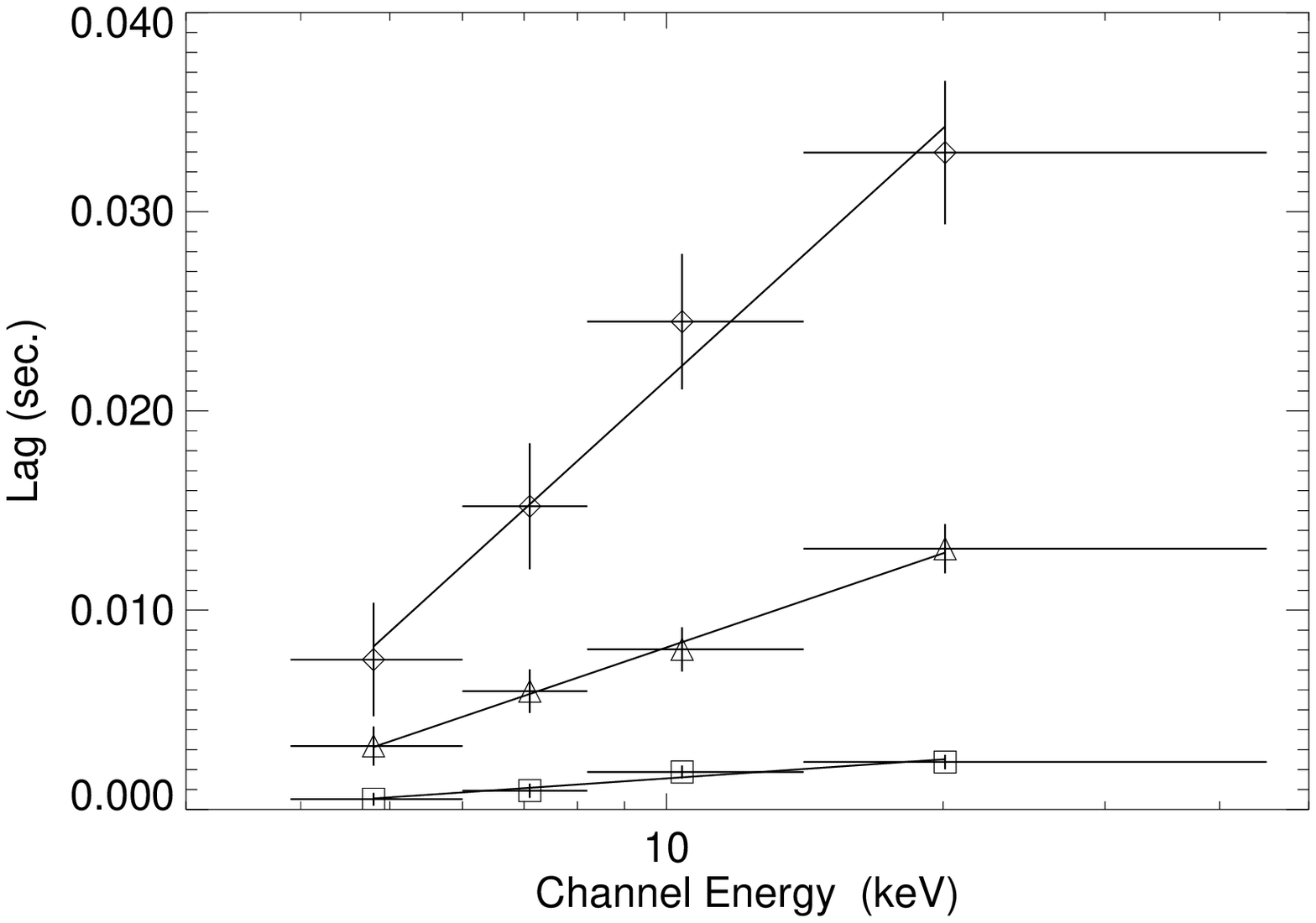}
\end{center}
\caption{Left: PSD of \cyg\ fit with five Lorentzian features.  The
four most prominent ones seem to be persistent in the hard state PSD,
and vary with spectral properties (see previous figure).  Middle: Time
lag between hard and soft X-ray variability for a hard state
observation of \cyg\ (see Nowak et al. 1999).  Note the peaks near the
PSD peak frequencies. Right: The energy dependence of the hard X-ray
variability lag for three different PSD frequencies (see Nowak et
al. 1999).  }
\label{fig:cygpsd}
\end{figure}

As we (and others) have shown, the Power Spectral Densities (PSD) of
the X-ray variability of \cyg\ can be well-described as the sum of
multiple, broad Lorentzian components \cite{nowak:00a}, with four
components dominating the fits (Fig.~\ref{fig:cygpsd}).  The peak
frequencies of these features are well-correlated with spectral
hardness, with harder spectra corresponding to lower frequencies
\cite{pottschmidt:02a}.  This is what one expects in sphere+disk
models if the frequencies are indicative of characteristic disk time
scales at the radius of the transition region between corona and disk,
and this transition radius moves outward as the source becomes fainter
and harder.  If this is the case, then the highest frequency component
at $\approx 40$\,Hz indicates that \emph{the transition region never
moves beyond $\approx 40$\,$GM/c^2$}.  (This is in contrast to many of
the larger values found referenced for ADAF models. As such, these
models have greatly scaled down their hypothesized coronal
region sizes in recent years.)

Often neglected is the fact that while the characteristic variability
time scales increase with hardness, the characteristic time lags
between soft and hard X-ray variability \emph{decrease}
\cite{pottschmidt:02a}.  This has no obvious explanation in the
scenario where the transition radius grows as the source fades into
quiescence.  Furthermore, the time lag between soft and hard X-ray
variability seems to be composed of multiple components at different
frequencies, very possibly associated with the individual Lorentzian
components in the PSD (Fig.~\ref{fig:cygpsd},
\cite{nowak:99a,nowak:00a}).  The time lag we observe may be in
reality a composite of time lags \emph{and leads} from independent
components, and this possibility is absent in most models of these
data.  Additionally, it is known that the time lag has a logarithmic
dependence upon energy, and in the past this had been used to argue
for Comptonization models (where logarithmic energy dependences are
naturally expected); however, given the \emph{extremely} long time
scale of the lags relative to dynamical time scales, the inferred
coronal size is unreasonably large \cite{nowak:99b}. The large
magnitude of the time lag must be incorporated in any model.

\begin{figure}
\begin{center}
\includegraphics[width=0.48\textwidth]{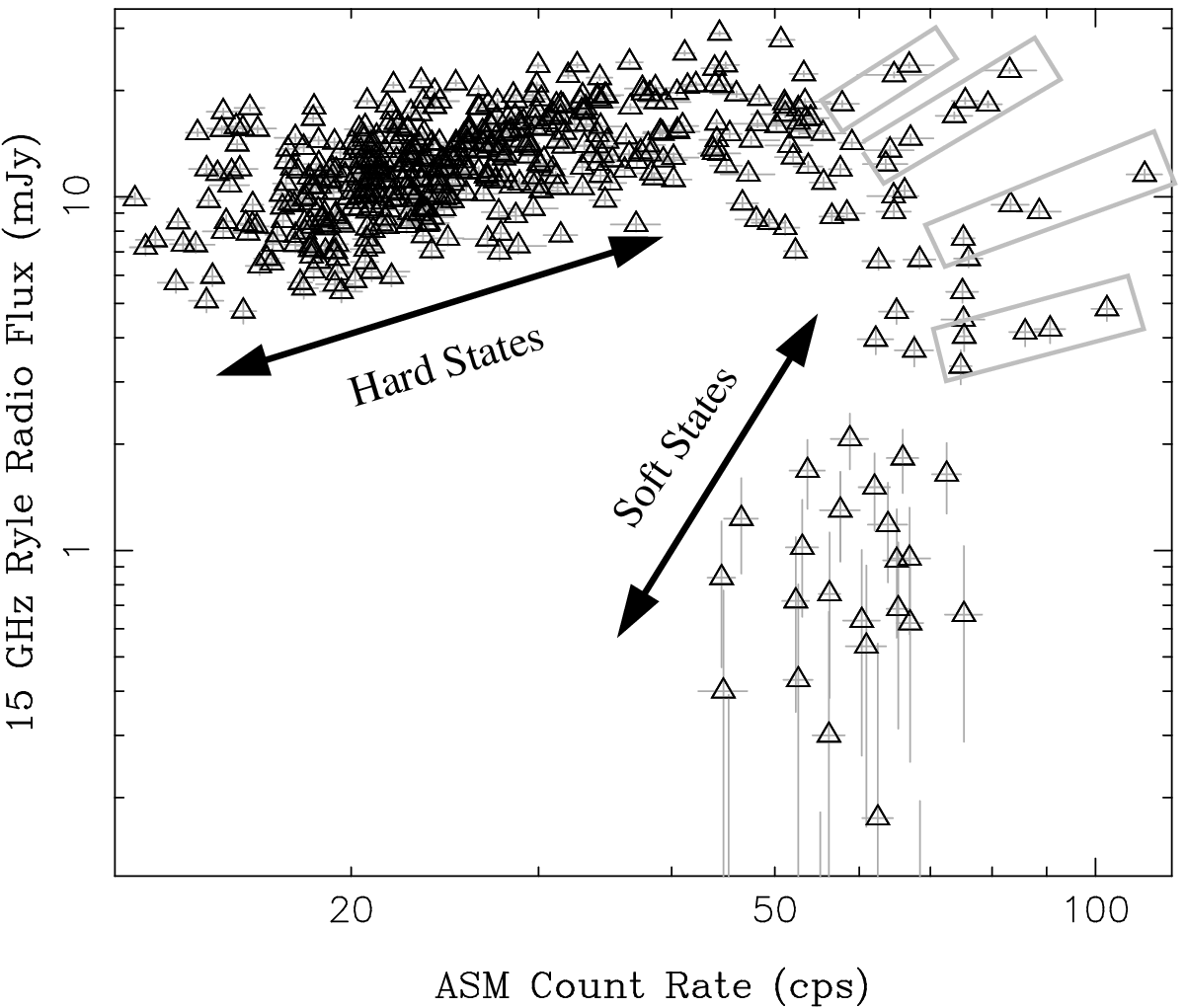}
\includegraphics[width=0.42\textwidth]{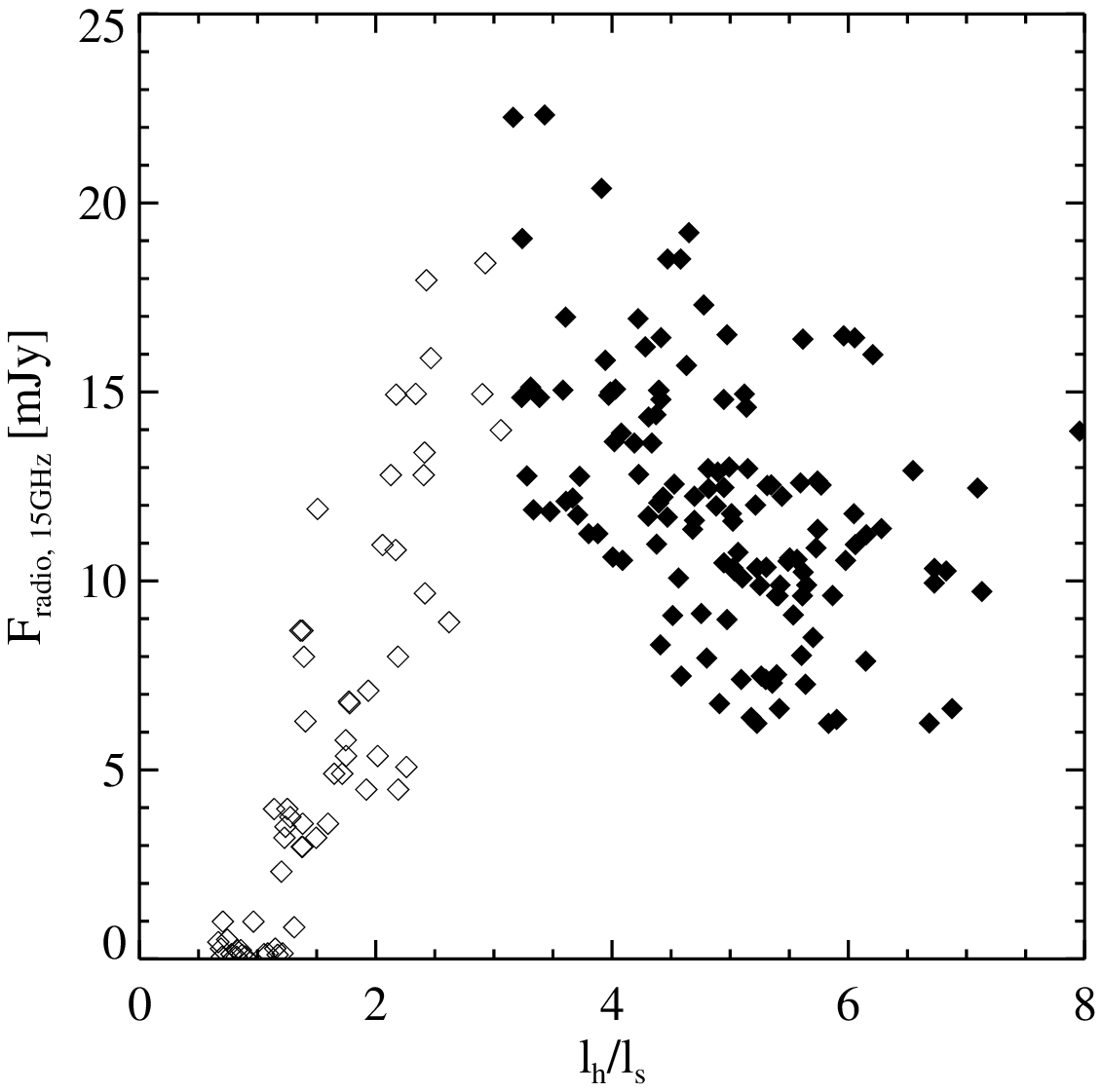}
\end{center}
\caption{Left: 15 GHz radio flux, measured with the Ryle radio
telescope, vs. daily average \textsl{ASM} count rate for \cyg(see
Nowak et al. 2005).  Boxes correspond to ``failed state transitions''.
Right: Ryle radio flux vs. coronal compactness for our pointed
observations of \cyg.}
\label{fig:cygrad}
\end{figure}

Finally, in all of the above we have not addressed any of the radio
data.  Nearly all of our \cyg\ observations have simultaneous 15\,GHz
radio data obtained with the Ryle radio telescope.  As shown in
Fig.~\ref{fig:cygrad}, although the radio can fade with increasing
flux/decreasing hardness, it rarely fully disappears in \cyg.  This is
partly what we mean by saying that state transitions are not distinct
in \cyg\ -- there appears to be a continuum of observed properties
between the hardest and softest spectra.  In fact, there are only two
properties where the \cyg\ state transitions do seem sharply defined.
First, variability time lags seems to greatly lengthen in state
transitions and failed state transitions \cite{pottschmidt:02a}.
Second, as shown in Fig.~\ref{fig:cygrad} (see also \cite{wilms:06a}),
the slope of the radio/X-ray hardness correlation changes from one
state to the other (although this figure, too, straightens out if one
plots radio vs. hard X-ray flux; \cite{nowak:05a}).  In nearly all
Compton models to date, the radio data, which clearly is an important
aspect of the source properties, is an {\sl ad hoc} add-on, with only
vague arguments as to its correlation with the X-ray.

\begin{figure}
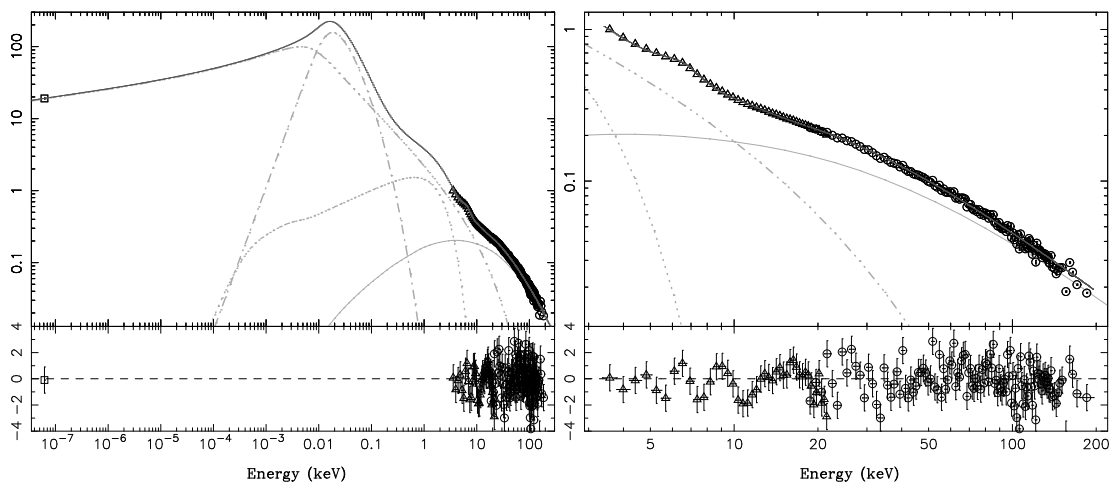

\begin{center}
\includegraphics[width=0.48\textwidth]{cyg_4099_19_all_II.ps}
\includegraphics[width=0.48\textwidth]{cyg_4099_19_xray_II.ps}
\end{center}
\caption{Jet model fit to radio and X-ray hard state data of \cyg\
(see Markoff, Nowak, \& Wilms 2005).  The model components include
disk emission, synchrotron, synchrotron self-Compton, and
Comptonization of disk photons. In the X-ray regime, the models work
equally as well as traditional corona models.}
\label{fig:cygjet}
\end{figure}

The need to self-consistently fit the radio and X-ray data together
has given rise to jet models of BHC \cite{markoff:05a}.  The
earliest versions of these models were dominated solely by synchrotron
radiation in the X-ray regime; however, for over three years now,
these models have included synchrotron, synchrotron self-Compton
(SSC), disk photons, and Comptonization of disk photons.  Other
reviews will go into these models in more detail; however, we will
point out two salient facts.  First, as shown in
Fig.~\ref{fig:cygjet}, these models not only fit the radio data, they
also simultaneously fit the \rxte\ X-ray data equally well as the
Comptonization models.  Second, like the Comptonization models, the
X-ray is dominated by two broad-band continuum components that in part
lead to the 10\,keV power law break.  As opposed to being disk plus
Comptonization components, these two components instead are
synchrotron and SSC.  (The upper break is also partly influenced by a
Comptonized disk component).  Again, as there is a broad Fe line,
reflection \emph{must} be a part of these spectra.  The specific
reflection and line parameters one fits, however, will change
depending upon whether one assumes a Compton corona or X-ray emitting
jet model.

\section{\fu}

\begin{figure}
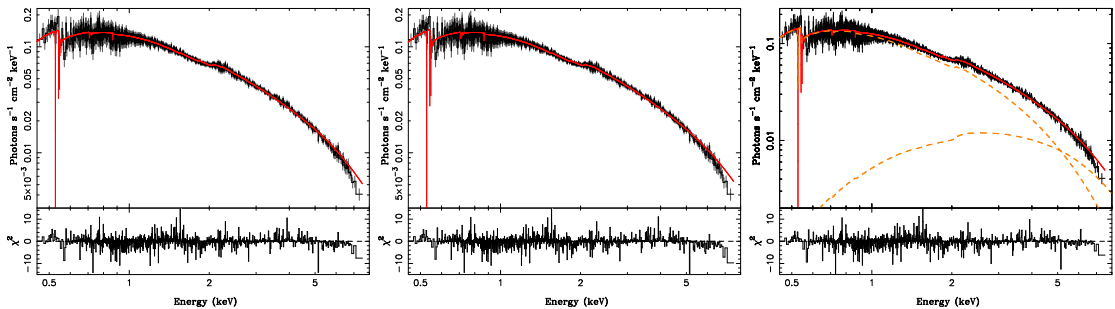

\begin{center}
\includegraphics[width=0.32\textwidth]{meg_kerrbb_thawed_hd_fixed_dbh.ps}
\includegraphics[width=0.32\textwidth]{meg_kerrbb_thawed_hd_fixed_dbh_a0.ps}
\includegraphics[width=0.32\textwidth]{meg_comptt.ps}
\end{center}
\caption{Left: The {\tt kerrbb} model fit to the combined \meg\ data
of \fu.  System distance, inclination, and mass were fixed (see text),
but spectral hardening factor and black hole spin were left
free. Middle: same data and model as on the left, except that now the
black hole spin is fixed to $a^* = 0$. Right: A disk plus
Comptonization model (see text) fit to the combined \meg\ data.
Dashed lines show the individual model components.
\label{fig:kerrbb}}
\end{figure}  

Contrary to \cyg, \fu\ is a source that seems to spend nearly all of
its time in a spectrally soft (i.e., ``high'') state. It shows
evidence of a hard tail only at its highest luminosities
\cite{wijnands:02c}, thus \fu\ cannot be used as a test of jet
models. Instead, \fu\ provides us with an opportunity to test recent
sophisticated disk atmosphere models that incorporate the effects of
black hole spin on the spectra (e.g., \cite{li:05a}). If such models
truly can unambiguously measure black hole spin as has been
hypothesized \cite{shafee:06a}, then they can be applied to systems
that transit between hard and soft states to determine whether rapid
spin is indeed a necessary component of jet production (thus testing
the \emph{Fender Conjecture}). We have obtained a simultaneous
\chandra/\rxte\ observation, and furthermore have analyzed all \rxte\
observations from the archive. Below we apply the {\tt kerrbb} disk
model \cite{li:05a}, which includes spin, to these spectra.

The \chandra\ spectra are shown in Fig.~\ref{fig:kerrbb}.  Owing to
the high resolution and the 0.4\,keV lower bound of our \chandra\
gratings observation, we can accurately measure the neutral column
($10^{21}~{\rm cm}^{-2}$) in front of the source, which removes one
potential source of ambiguity in the models.  Although predominantly
being merely phenomenological, the two-parameter {\tt diskbb} model
fits the data \emph{very} well.  Thus, similar to our point about
broken power law fits to \cyg, any model that attempts to describe the
data with more than two parameters is likely over-determined.  The
{\tt kerrbb} model has seven: system mass ($M$), accretion rate ($\dot
M$), distance ($D$), inclination ($\theta$), spectral hardening factor
(\fc), torque parameter ($\eta$), and dimensionless black hole spin
(\as).

\begin{figure}
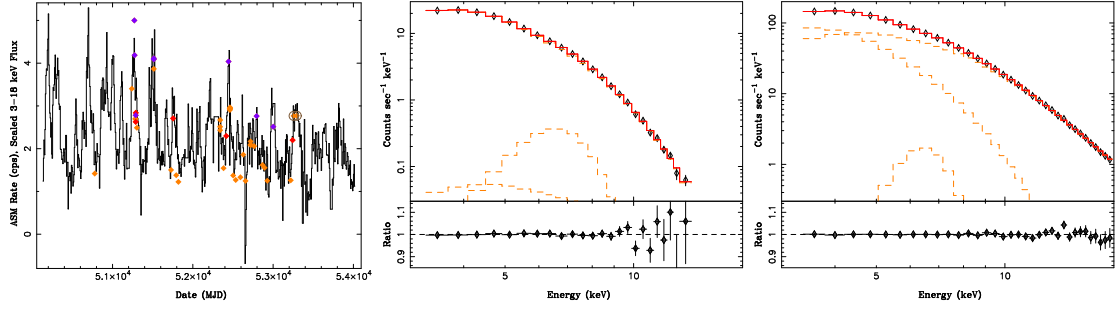

\begin{center}
\includegraphics[width=0.32\textwidth]{lightcurve.ps}
\includegraphics[width=0.32\textwidth]{rxte_70054_01_11_kIII.ps}
\includegraphics[width=0.32\textwidth]{rxte_40044_01_02_kIII.ps}
\end{center}
\caption{Left: \textsl{All Sky Monitor} lightcurve for \fu\, with scaled
3-18\,keV \pca\ flux from pointed observations overlaid.  \rxte\
observations simultaneous with \chandra\ and \xmm\ are circled.
Middle, Right: Count rate spectra for the softest/faintest (middle) and
hardest/brightest (right) of the \rxte\ observations, fit with the
{\tt kerrbb+comptt+gaussian} model.  Dashed lines show the individual
model components folded through the detector response.
\label{fig:lightcurve}}
\end{figure}

In practical application of the {\tt kerrbb} model, it is hoped that
$M$, $D$, and $\theta$ can be determined via other observations and
that \fc\ and $\eta$ can be determined from theoretical
considerations, which would leave only $\dot M$ and \as\ as fit
parameters.  The complex optical lightcurve of \fu\ implies an
inclination of $\sim 75^\circ$ \cite{hakala:99a}; however, the system
mass and distance are completely unknown. The lowest luminosity \rxte\
observations \emph{do not} transit to the hard state; although, an
upturn in the fitted disk normalization from {\tt diskbb+powerlaw}
models (Fig.~\ref{fig:lightcurve}) indicates they may be very near the
transition, expected to be at $\approx 3\%~L_{\rm Edd}$
\cite{maccarone:03a}.  Thus the least massive/closest \fu\ could be is
3\,M$_\odot$ at 10\,kpc (used in all figures shown here).  The
expected distance then scales as the square root of the mass (e.g.,
16\,M$_\odot$ at 23\,kpc).

Simple {\tt diskbb} fits to the spectra yield very high temperatures
($kT \sim 1.7$\,keV), and low normalizations ($\sim 8$).  \emph{If one
fixes \fc\ in the {\tt kerrbb} model}, these {\tt diskbb} parameters
are only achievable with a combination of large distance, high
accretion rate, \emph{and} high spin.  (High spin becomes more
crucially needed if one increases the mass of and distance to \fu.)  A
high spin model (with \fc $\sim 1.1$) is shown in
Fig.~\ref{fig:kerrbb}.  Note, however, that we can find a nearly
equally good fit with \as$=0$ if we allow \fc $=3.3$
(Fig.~\ref{fig:kerrbb}).  Thus, \emph{at a minimum}, one must be
absolutely convinced that there is a strong theoretical motivation for
specific values of \fc\ and $\eta$ if disk models are to be used to
`measure' spin.

\begin{figure}
\begin{center}
\includegraphics[width=0.95\textwidth]{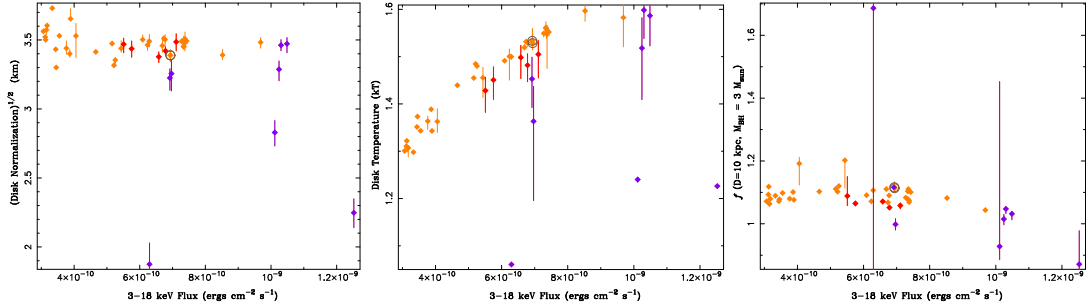}
\end{center}
\caption{Left: Square root of disk normalization vs. \pca\ flux for
{\tt diskbb+powerlaw} fits to \rxte\ observations of \fu.  Middle:
disk temperature vs. \pca\ flux for {\tt diskbb+comptt} fits to \fu.
Right: Spectral hardening factor vs. \pca\ flux for {\tt
kerrbb+comptt} fits to \fu.  (For the latter, we assume a black hole
mass of 3\,$M_\odot$, a distance of 10\,kpc, and an inclination of
75$^\circ$.)
\label{fig:rxtefig}}
\end{figure}  

As opposed to the \chandra\ data, where \as $\approx 1$ and \as $=0$
are virtually indistinguishable, the very high statistics \rxte\ data
seem to require \as $\approx 1$ (and find \fc $\approx 1.1$).  If one
sets the color correction factor to the `theoretically preferred'
value of \fc $= 1.7$, and requires that the faintest observations have
a luminosity of $\approx 3\%~L_{\rm Edd}$, then the \rxte\ data
\emph{require} \as $\approx 1$, $D \approx 23$\,kpc, $M \approx
16~M_\odot$.  (The degeneracy among all these models is partly
indicative of the fact that the produced spectra are rather smooth,
with no specific sharp features unique to \as $=1$, for example.)
\emph{If} such models can usefully be used to measure \as, then a
prediction of the {\tt kerrbb} model, based on the \rxte\ data, is
that future observations should find \fu\ to be approximately
16\,M$_\odot$ and near 23\,kpc, i.e., well into the galactic halo.

A further serious issue in applying the {\tt kerrbb} model arises in
looking at both the \chandra\ and \rxte\ data.  If one allows a
hardening due to Comptonization (here, modeled as the addition of the
{\tt comptt} model with a fixed coronal temperature and seed photon
temperature frozen to the best-fit {\tt diskbb} temperature), one
achieves an even better fit to the \chandra\ data than either the {\tt
diskbb} or {\tt kerrbb} models alone.  Furthermore, the best fit {\tt
diskbb} temperature decreases to 1.3\,keV, and the best fit
normalization increases to 15.  That is, the very parameters driving
the need for high spin are fundamentally altered to values indicative
of much lower spin.  The fit to the \chandra\ data essentially leaves
the spin parameter unconstrained in a systematic, rather than a
statistical, sense.

\rxte\ spectra, however, still require a high spin parameter.  As
shown in Fig.~\ref{fig:rxtefig}, there are periods when the disk
temperature and normalization drop dramatically, and the spectrum
becomes dominated by the Compton component
(Fig.~\ref{fig:lightcurve}). These correspond to \fu\ entering the
``very high'' or ``steep power law'' state.  Otherwise, flux
variations are predominantly driven by disk temperature changes in
{\tt diskbb} models, or accretion rate changes in {\tt kerrbb} models.
Note in Figs.~\ref{fig:kerrbb} and \ref{fig:lightcurve} that even when
dominated by the Compton component, the \fu\ spectrum is very soft,
and only differentiated from the disk spectrum by a decrease in
spectral curvature\footnote{Note that we also include a weak Fe line,
likely due to galactic emission, in the fits.  When the {\tt kerrbb}
model has been fit to other sources, researchers have often required
stronger lines, smeared edges, and even more prominent power law
components than required here, adding greatly to systematic
uncertainties in the use of these disk models \cite{shafee:06a}.}.

In Fig.~\ref{fig:rxtefig} we also see that as the flux increases, the
fitted spectral hardening factor, \fc, decreases.  That is,
Comptonization is acting to harden the disk spectrum, rather than it
being a necessary parameter in the disk atmosphere model itself.  As
the flux decreases, the hardening factor, \fc, smoothly increases,
perhaps asymptoting to a `pure disk' value.  Given the fact even the
Compton-dominated spectra are so soft, and there seems to be a
continuous evolution of \fc, can one really be sure that the lowest
flux spectra are indeed ``pure'' disk spectra?  How does one know that
there isn't a ``residual corona'' that merely mimics the effects of
high spin in the disk model?  For these reasons, I am very skeptical
that such models will ever usefully ``measure'' spin.  (However, if
independent observation determines that \fu\ is indeed a 16\,M$_\odot$
black hole at 23\,kpc, I may become a believer $\ldots$)

\section{\gx}

\begin{figure}
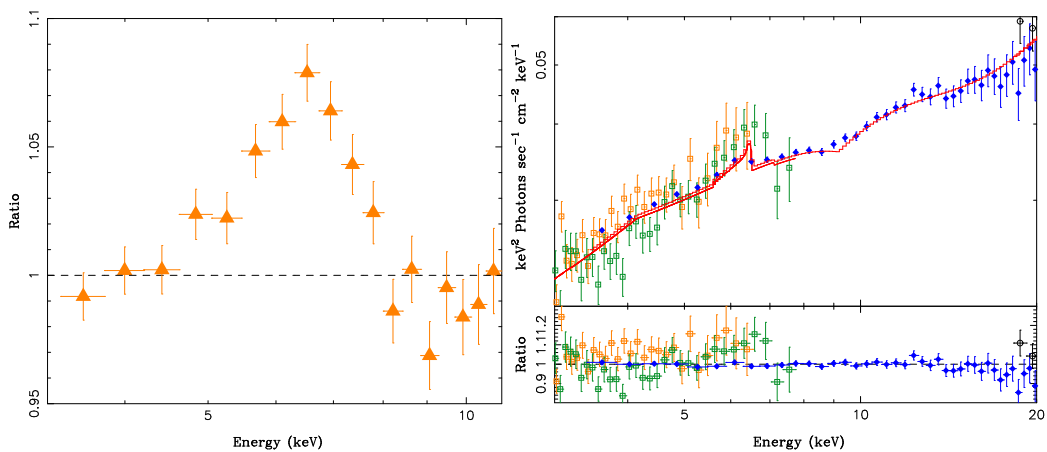

\begin{center}
\includegraphics[width=0.435\textwidth]{line_res.ps}
\includegraphics[width=0.47\textwidth]{diskline.ps}
\end{center}
\caption{Left: Fe line residuals from an \rxte\ observation of a \gx\
hard state as it enters quiescence.  The line remains broad, contrary
to the expectation of ADAF models, despite the low flux (see Nowak,
Wilms, \& Dove 2002).  Right: An even fainter hard state \rxte/\chandra\ observation
of \gx, potentially showing a broad line (Nowak et al., in prep.)
\label{fig:lineres}}
\end{figure}  

\begin{figure}
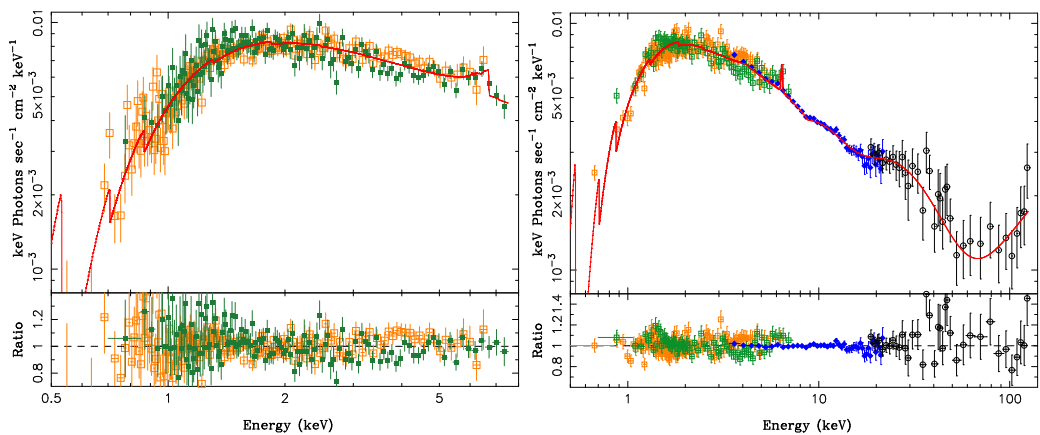

\begin{center}
\includegraphics[width=0.45\textwidth]{diskbb_power_chandra_only.ps}
\includegraphics[width=0.45\textwidth]{very_funky.ps}
\end{center}
\caption{Left: The broad-band \chandra\ spectra.  The model is
consistent with a disk whose inner edge extends to the marginally
stable orbit, plus a broad line.  Right: The broad-band
\chandra/\rxte\ spectra.  The observed (multiple humped) curvature,
and flat high energy tail, are real.  This is not your grandmother's
broken power law, and does not fit either a simple corona or jet model
(Nowak et al. in prep.)
\label{fig:funky}}
\end{figure}  

\gx\ has been a very important source in establishing the relationship
between hard X-ray flux and steady radio jet activity
\cite{hanni:98a,fender:99b,corbel:00a,corbel:03a}.  We have
extensively studied the spectra and variability properties of this
source ourselves, and have successfully fit both Comptonization and
jet (and, of course, broken power law) models to the hard state
spectra
\cite{wilms:99aa,nowak:99c,nowak:00a,nowak:02a,nowak:05a,markoff:05a}.
Here, I will only briefly touch upon two faint, hard state
observations of \gx.

The `sphere+disk' Comptonization models posit a transition radius
between inner corona and outer disk, and likewise ADAF models
similarly posit a transition radius from outer, efficient flow to
inner, inefficient flow.  The former models do not set a specific
radius for this transition (although as discussed earlier, timing data
of \cyg\ indicate that it cannot occur at $R > 40~GM/c^2$).  The ADAF
models have traditionally posited greater transition radii, with the
radius possibly increasing as the source falls further into
quiescence.  Thus, especially for the latter models, we do not expect
to see broad Fe lines persist into quiescence. \gx\ shows at least
one, possibly two, counter-examples to that ADAF expectation.

As discussed in \cite{nowak:02a}, and shown in Fig.~\ref{fig:lineres},
the 1999 fade into quiescence saw the Fe line remain strong and
persistently broad.  Despite the fact that the source was
approximately a factor of 10 lower in flux from its hard state
transition level, the best fits indicated that the inner edge of the
line emission region was consistent with the marginally stable orbit.
This is completely inconsistent with ADAF theory.  

More recently, during the 2005 fade into quiescence, we obtained a
simultaneous \chandra/\rxte\ observation at an even fainter hard state
flux level.  As shown in Fig.~\ref{fig:lineres}, these data are
\emph{also} consistent with a relativistically broadened line, counter
to the expectations of ADAF models (Nowak et al., in prep.).  Here,
however, owing to the faint nature of these spectra, the fits are more
uncertain (galactic ridge emission must be carefully subtracted from
the \rxte\ spectra); however, the broad line does seem to be preferred
in the \chandra\ data.  Likewise, if one fits a {\tt diskbb+powerlaw}
model to solely the \chandra\ data, the fitted disk also prefers an
inner radius consistent with the marginally stable orbit, rather than
the much larger disk radii often fit in hard state sources
(Fig.~\ref{fig:funky}).

The 1999 fade into quiescence was `regular' in all respects, and was
instrumental in defining the the $F_{radio} \propto F_{X-ray}^{0.7}$
correlation between radio and X-ray fluxes in hard state BHC
\cite{corbel:03a}.  The presence of the broad line in those data is also
unambiguous, and thus is challenging for ADAF theory.  On the other
hand, the 2005 fade into quiescence was very unusual.  As discussed by
S. Corbel in these proceedings, a few weeks before the \chandra/\rxte\
observation, \gx\ `fell off' the usual radio/X-ray correlation, with
the radio decreasing \emph{much} more rapidly.  Relative to the X-ray
flux (which itself was very low, given that the observations occurred
only three weeks after the hard state transition), the radio was
approximately a factor of ten to faint for the usual correlation.

Furthermore, the broad-band X-ray spectra themselves, as shown in
Fig.~\ref{fig:funky}, are highly unusual.  The data show a great deal
of curvature and spectral breaks, and there is a very spectrally flat
high energy tail.  \emph{All} of these features are real, and are
\emph{not} artifacts of the unfolding process.  (Remember, we're not
using \textsl{XSPEC} $\ldots$).  We have also been very careful in the
background subtraction of the galactic ridge emission.  These data
simply represent a very unusual, radio-weak hard state spectrum.
Currently, we have no good model with either typical corona or typical
jet models.  (Broken power law models don't even work well here.)  Why
\gx\ failed to act as a `proper' microquasar in this particular
outburst decay remains a mystery.

\acknowledgments I'd like to thank my coauthors and collaborators on
the above cited projects: Stephane Corbel, James Dove, Thomas
Gleissner, Adrienne Juett, Sera Markoff, Guy Pooley, Katja
Pottschmidt, \& J\"orn Wilms.  Intelligent insight is largely thanks
to them, ranting is wholy my own. If there were less of the latter,
and this proceedings weren't already a month late, many of them would
be coauthors here.  I would like to thank the organizers for inviting
me, and NASA Grant SV3-73016 for funding this work.


\end{document}